\documentclass[]{PoS}
\usepackage{amsfonts,amsmath}

\newcommand{\psibar}{\ensuremath{\bar\psi}}
\newcommand{\chibar}{\ensuremath{\bar\chi}}
\newcommand{\pbp}{\ensuremath{\psibar\psi}}
\newcommand{\pbpvac}{\ensuremath{\langle \psibar\psi \rangle}}
\newcommand{\vev}[1]{\ensuremath{\left<#1\right>}}
\newcommand{\vevsub}[1]{\ensuremath{\left<#1\right>_{\mathrm{sub}}}}

\newcommand{\Real}{\ensuremath{\mathrm{Re}}}

\newcommand{\Trace}{\ensuremath{\mathrm{Tr}}}

\newcommand{\Eq}[1]{Eq.\,(\ref{#1})}

\newcommand{\Fig}[1]{Fig.\,\ref{#1}}

\newcommand{\Cite}[1]{Ref.\ \cite{#1}}

\newcommand{\tab}[1]{Table \ref{#1}}

\newcommand{\eg}{e.\ g.\ }

\newcommand{\links}{\textbf{Left}: }

\newcommand{\rechts}{\textbf{Right}: }

\title{\begin{center} Towards thermodynamics with $N_f=2+1+1$ \\twisted mass quarks \end{center}}

\ShortTitle{}

\author{{\textbf tmfT Collaboration:}}
\author{\speaker{F. Burger}, G. Hotzel, M. M{\"u}ller-Preussker\\
        Humboldt-Universit\"at zu Berlin, Institut f\"ur Physik, 12489 Berlin, Germany
}
\author{E.-M. Ilgenfritz\\
       Joint Institute for Nuclear Research, VBLHEP, 141980 Dubna, Russia}
\author{M. P. Lombardo \\
       Laboratori Nazionali di Frascati, INFN, 100044 Frascati, Roma, Italy}

\leftline{\parbox{11cm}{\large\rm HU-EP-13/55, SFB/CPP-13-85}}
\vspace*{-0.5cm}

\abstract{We present preliminary results achieved within a recently started project dealing with QCD thermodynamics in the presence of a fully 
dynamical second quark family. We are employing the Wilson twisted mass discretization. To reduce the amount of zero temperature
simulations and the cost of analysis we have chosen the fixed-scale approach. We show a variety of basic thermodynamic observables for 
temperatures ranging from 158 to 633 MeV. 
Simulations were performed for three lattice spacings below $0.1$ fm each and at a single value of the 
pion mass which allows a comparison with previously obtained $N_f=2$ results. We determine the chiral crossover temperature from 
the bare chiral susceptibility and show results for the 
gauge part of the trace anomaly.
}

\FullConference{31st International Symposium on Lattice Field Theory - LATTICE 2013\\
		July 29 - August 3, 2013\\
		Mainz, Germany}

\begin{document}

\section{Introduction}
The LHC is able to study the quark gluon plasma (QGP) up to temperatures of six times the transition temperature, where
one would expect effects of the charm quark being non-negligible. While in the crossover region 
below temperatures of $200 \mathrm{~MeV}$ the charm might well be neglected it will contribute to 
the equation of state (EoS) at higher temperatures. The EoS is of central interest as a necessary input 
for the hydrodynamical description of the evolution of the plasma created in heavy-ion collisions. Lattice calculations are well able to provide the EoS, 
however only very few calculations including dynamical charm 
have been performed so far \cite{Bazavov:2012kf,Ratti:2013uta}.
Both these studies have used the staggered fermion discretization. No four-flavor results exist so far that use a Wilson-type quark action.

The tmfT collaboration has been mostly studying two-flavor QCD \cite{Burger:2011zc} with the focus on the order and universality class of the 
chiral limit of the phase transition. In a contribution to the Lattice Field Theory Symposium 2013 we have reported about 
early and preliminary results of a new project that addresses the thermodynamics in the presence of the second quark family.
We have chosen to scan the temperature $T= 1/(N_\tau a)$ by a simultaneous
study of lattices with a varying number of temporal lattice steps in imaginary time
direction $N_\tau$. This approach has been advocated in Ref.~\cite{Umeda:2012er}.

Adopting the same lattice discretization and setup that is used by the European Twisted Mass
Collaboration (ETMC) for their $N_f=2+1+1$ simulations \cite{Baron:2009zq,Baron:2011sf} at zero temperature, we can take advantage of their large 
set of $T=0$ gauge field ensembles as well as the scale setting already done.
For the present study we have taken the values $a = 0.0863(4) \mathrm{~fm},  ~0.0779(4) \mathrm{~fm}$ and $0.0607(2) \mathrm{~fm}$ from \cite{Baron:2011sf}.

\vspace{-0.3cm}
\section{Lattice Action and Simulation Parameters}
In terms of 
the twisted fields $\chi_{l,h}  = \exp{(- i \pi \gamma_5 \tau^{3} / 4 )} \psi_{l,h}$ 
the light and heavy quark twisted mass action  
have the following form:
\begin{equation}
S^l_f[U,\chi_l,\chibar_l] = \sum_{x,y} \chibar_l(x) \left 
 [\delta_{x,y} -\kappa D_\mathrm{W}(x,y)[U] + 2 i \kappa a \mu \gamma_5 \delta_{x,y}  \tau^3 \right ]
 \chi_l(y) \;,
\label{tmaction}
\end{equation}
and similarly:
\begin{equation}
  S^h_f[U,\chi_h, \overline{\chi}_h] =\sum_{x,y} \overline{\chi}_h(x) \left[ \delta_{x,y} - \kappa D_W(x,y)[U] + 2 i \kappa \mu_{\sigma} \gamma_5 \delta_{x,y} \tau^1 + 2 \kappa \mu_{\delta} \delta_{x,y}
\tau^3 \right] \chi_h(y) \;.
\label{eq_heavyaction}
\end{equation}
For the gauge sector the Iwasaki action is used ($c_0 = 3.648$ and $c_1 = -0.331$):
\begin{equation}
S_g[U] = \beta  \Big{(} c_0 \sum_{P} \lbrack 1 - \frac{1}{3} \Real \Trace \left ( U_{P} \right ) \rbrack 
 + c_1 \sum_{R} \lbrack  1 - \frac{1}{3} \Real \Trace  \left ( U_{R} \right ) \rbrack \Big{)} \;.
\label{tlsym}
\end{equation}
The two sums extend over all possible plaquettes ($P$) and planar
rectangles ($R$), respectively.

For the generation of finite temperature gauge field configurations 
we have adopted the same bare parameter set as three of ETMC's gauge field ensembles labelled by
A60.24, B55.32 and D45.32 that correspond to $a ~\sim 0.086, 0.078$ and $0.061 \mathrm{~fm}$, respectively. The charged pion mass is tuned to $\sim 400 \mathrm{~MeV}$ for these ensembles. We refer to 
\Cite{Baron:2011sf} for all details. In \tab{tab_simpar} we summarize the so 
far available finite temperature ensembles, which have been generated using 
the tmLQCD code package \cite{tmLQCD}.
\begin{table}
{\small
 \begin{center}
  \begin{tabular}{c|c|c|c}
  $T=0$ ensembles  & $N_\sigma$ & $N_\tau$  & $T [\mathrm{~MeV}]$\\
  \hline \hline
  A60.24  &  24 &  12,10,8,6,4 & 190 - 572 \\
          &  32 &  14     & 163       \\
  \hline
  B55.32  &  32 &  16,14,12,10,8,6,4 & 158 - 633 \\
  \hline
  D45.32  &  32 &  8,6  & 406, 542 \\
  \end{tabular}
\end{center}
}
\caption{Finite temperature simulation parameters and corresponding $T=0$ ensembles. See \Cite{Baron:2011sf} for 
the corresponding sets of bare parameter values.}
 \label{tab_simpar}
\end{table}

\vspace{-0.3cm}
\section{Observables in the Crossover Region}
\begin{figure}[htb] 
{\centering
\hfill
\includegraphics[height=5cm]{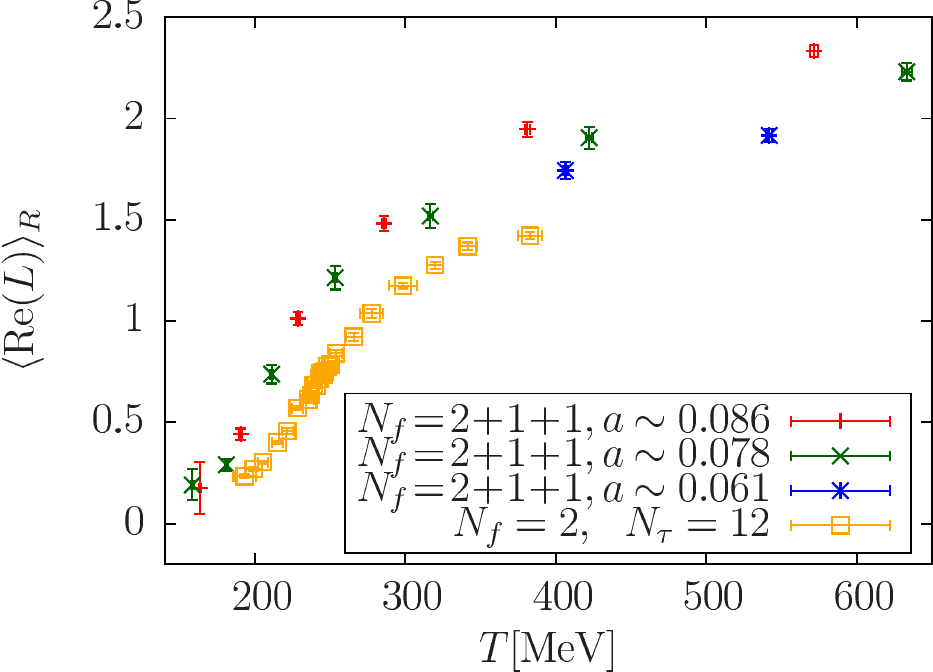}\hfill
\includegraphics[height=5cm]{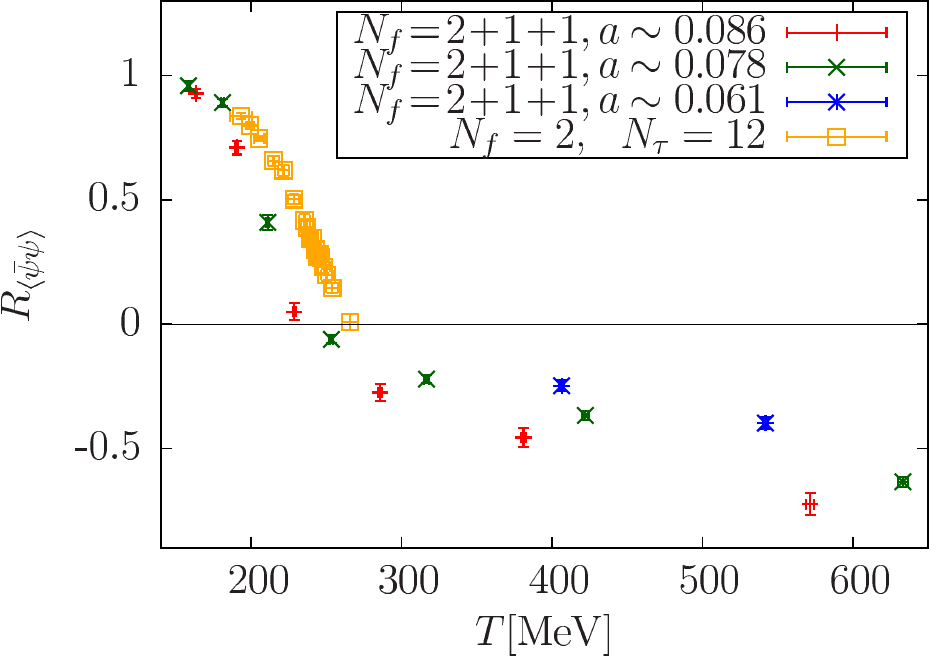}\hfill
}
 \caption[]{\links The renormalized Polyakov loop. The transition region is  so far only covered entirely by the two sets with coarsest lattice spacings.
\rechts The renormalized light chiral condensate according to \Eq{eq_pbpratio}.}
\label{fig_poly_pbpold}
\end{figure}

The transition from the hadronic phase to the QGP can be detected by studying suitable 
observables as \eg the Polyakov loop and the light chiral condensate.
However, since the corresponding unrenormalized quantities do not show a very pronounced temperature dependence, 
the renormalization of these observables is necessary. In the same way as in our previous studies with $N_f=2$, the Polyakov loop 
is renormalized with a renormalization factor evaluated from the static $\bar Q Q$ potential at temperature $T = 0$ and distance $r_0$:
\begin{equation}
 \vev{\Real(L)}_R = \vev{\Real(L)} \exp{(V(r_0)/2T)} \;.
 \label{eq_polyren}
\end{equation}
It is shown in the left panel of \Fig{fig_poly_pbpold} and seen to rise monotonously in the crossover region. For comparison we show in the same figure also our
two-flavor data obtained at the same value of the pion mass. For large temperatures ($\ge 350 \mathrm{~MeV}$)
a remaining cutoff effect is visible by a tendency of the data to attain lower values with decreasing lattice spacing. However, around the inflection 
point and thus in the vicinity of the crossover, the data from our two coarsest lattice spacings agree with each other. Thus we conclude that for $N_\tau \ge 8$ 
cutoff effects are small. From the overall view we note that with inclusion of the dynamical second quark generation the curve is shifted towards smaller temperatures as 
compared to $N_f=2$.

The chiral condensate $\pbpvac$ is evaluated by means of the technique of noisy estimators using quark matrix inversions on 24 Gaussian noise vectors.
These as well as all $T=0$ fermion matrix inversions were calculated on local GPU clusters.
Its additive quadratic, mass dependent divergence ($\propto \frac{\mu}{a^2}$) can be removed by subtracting the corresponding condensate at the same mass for $T=0$~\cite{Burger:2011zc}:
\begin{equation}
  R_{\pbpvac} = \frac{\pbpvac^{T,\mu}_l - \pbpvac^{T=0,\mu}_l + \pbpvac^{T=0,\mu=0}_l}{\pbpvac^{T=0,\mu=0}_l}\,.
  \label{eq_pbpratio}
\end{equation}
The multiplicative renormalization factor in the denominator is obtained by extrapolating the zero temperature condensate
to the chiral limit. To this end we have used a linear ansatz in the light mass, which was sufficient and 
has yielded acceptable fits. For the coarsest lattice spacing the pion mass had to be restricted to 
values lower than 370 MeV to obtain an acceptable fit.
We added a trivial unity in \Eq{eq_pbpratio} in order to account for the presence of symmetry breaking in the hadronic phase.
The same procedure has been followed in the two-flavor case for which we show $R_{\pbpvac}$ for comparison in the right panel of 
\Fig{fig_poly_pbpold}. Similarly to the Polyakov loop we observe stronger lattice artefacts in our data above $T=350 \mathrm{~MeV}$. Whether these alone can be blamed for the drop of $R_{\pbpvac}$ below zero above the transition
(where it is expected to vanish) 
should be further studied when more data for our finest lattice spacing becomes available also at lower temperature. 
Note again that around the crossover lattice spacing artefacts seem to be small and that the 
$N_f=2+1+1$ curve is shifted towards lower values of the temperature as was observed for the renormalized Polyakov loop, too.

\vspace{-0.3cm}
\section{Pseudo-Critical Temperature}
Having now the strange quark at our disposal, another prescription using a suitable subtraction of the strange quark condensate 
involving the light and strange masses can be used to eliminate the divergence in the 
light condensate. This procedure is used in the literature \cite{Cheng:2007jq}:
\begin{equation}
  \Delta_{l,s} = \frac{\pbpvac_l - \frac{\mu_l}{\mu_s}\pbpvac_s}{\pbpvac^{T=0}_l - \frac{\mu_l}{\mu_s}\pbpvac^{T=0}_s}\,.
   \label{eq_pbpsub}
\end{equation}

\begin{figure}[htb] 
{\centering
\hfill
\includegraphics[height=5cm]{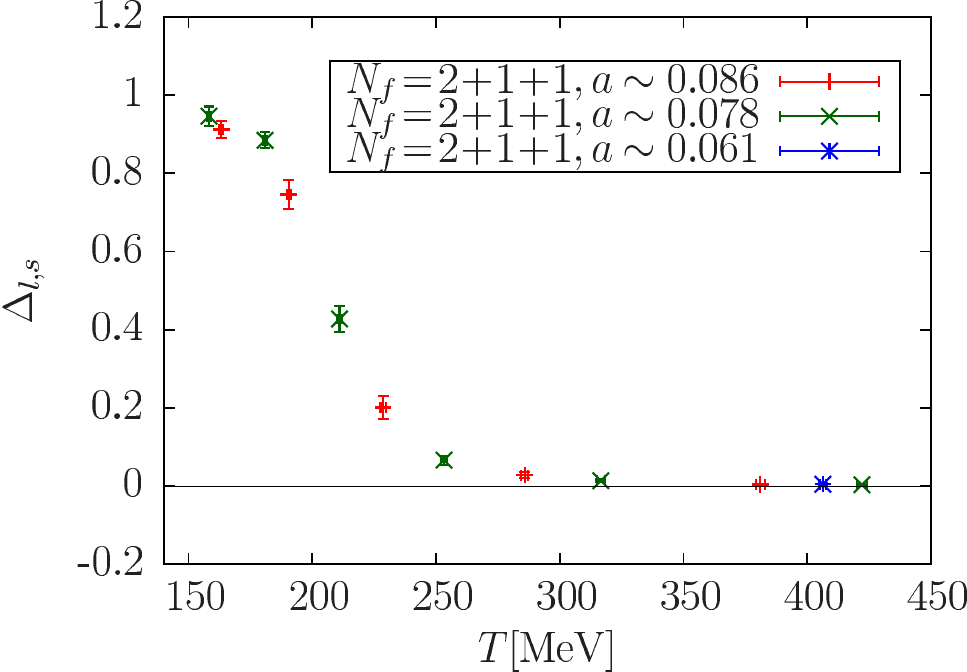}    \hfill
\includegraphics[height=5cm]{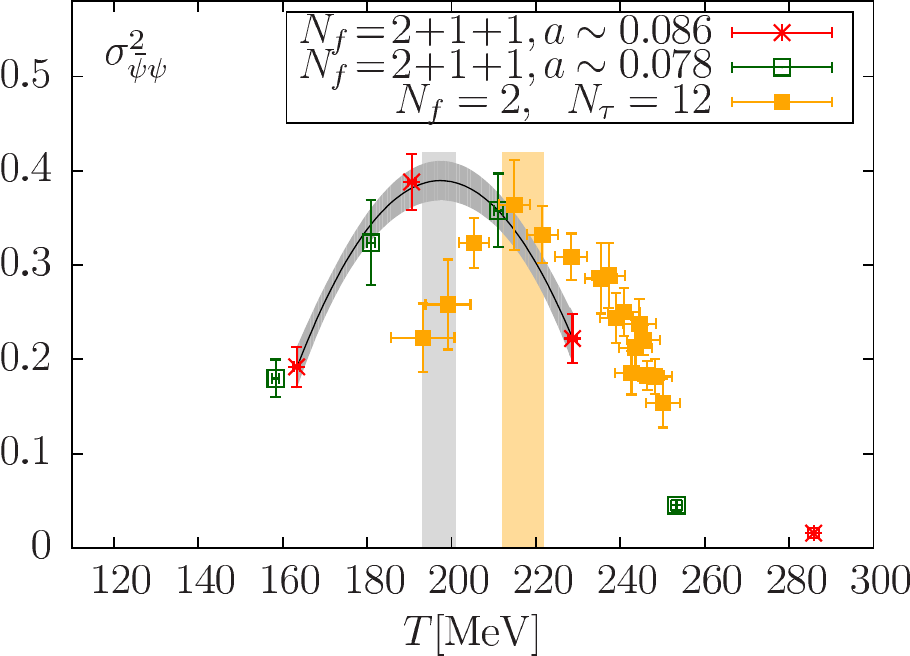} \hfill 
}
 \caption[]{\links The light chiral condensate according to \Eq{eq_pbpsub}. \rechts The disconnected part of the light
 chiral susceptibility with a polynomial fit to estimate the pseudo-critical temperature. The previously obtained $N_f=2$ data is included for comparison.
 The vertical bars indicate the estimated pseudo-critical temperatures.}
\label{fig_pbpnew}
\end{figure}
The strange quark condensate is obtained in the Osterwalder-Seiler setup \cite{Osterwalder:1977pc, Frezzotti:2004wz}, and the bare strange mass has been set as to reproduce the 
physical $\bar s \gamma_\mu s$-mass. 
As opposed to $\pbpvac_R$ of \Eq{eq_pbpratio}, the thus subtracted condensate shows no lattice spacing artefacts over the whole studied range of
temperatures and shows a smooth order parameter-like behavior.

In order to estimate the crossover temperature we have evaluated the disconnected part of the chiral susceptibility
\begin{equation}
\sigma^2_{\pbp} = V/T \left( \vev{(\pbp)^2} - \vev{\pbp}^2 \right)\;.
\label{eq_sigma2}
\end{equation}
Similar to the two-flavor case this quantity shows a maximum at the crossover temperature as can be appreciated in the right panel of \Fig{fig_pbpnew}.
The data of our two coarsest lattice spacings have been fitted with a polynomial ansatz in the temperature range
$160 \le T \le 240 \mathrm{~MeV}$ and gives a value of $T_\chi = 197.4(1.6)$. 
The corresponding fit curve is included in the figure. From varying the lower and upper bounds of fitting we estimate an 
additional systematic error ($\sim 3 \mathrm{~MeV}$) which we add in quadrature. 
For comparison we show the corresponding $N_f = 2$ data, which is located 
at larger temperatures consistent with the observations from the other observables. 
The result for $T_\chi$ is summarized in the following table:
\begin{center}
  \begin{tabular}{c|c|c}
      & $N_f = 2+1+1$                     & $N_f = 2$     \\
      \hline
      $T_\chi$ [MeV] & $197(4)$           & $217(5)$      \\ 
  \end{tabular}
  .
\end{center}

\vspace{-0.3cm}
\section{The Trace Anomaly}
\begin{figure}[htb] 
{\centering
\hfill
\includegraphics[height=5cm]{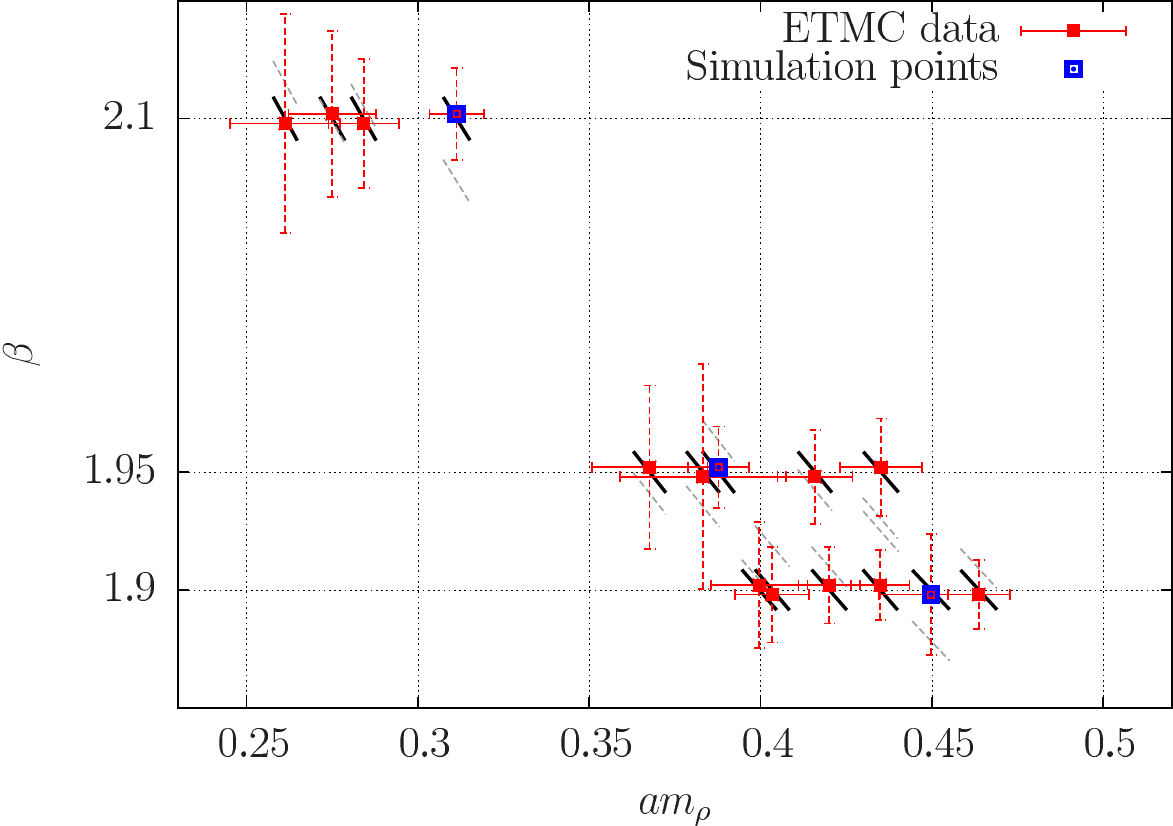} \hfill 
\includegraphics[height=5cm]{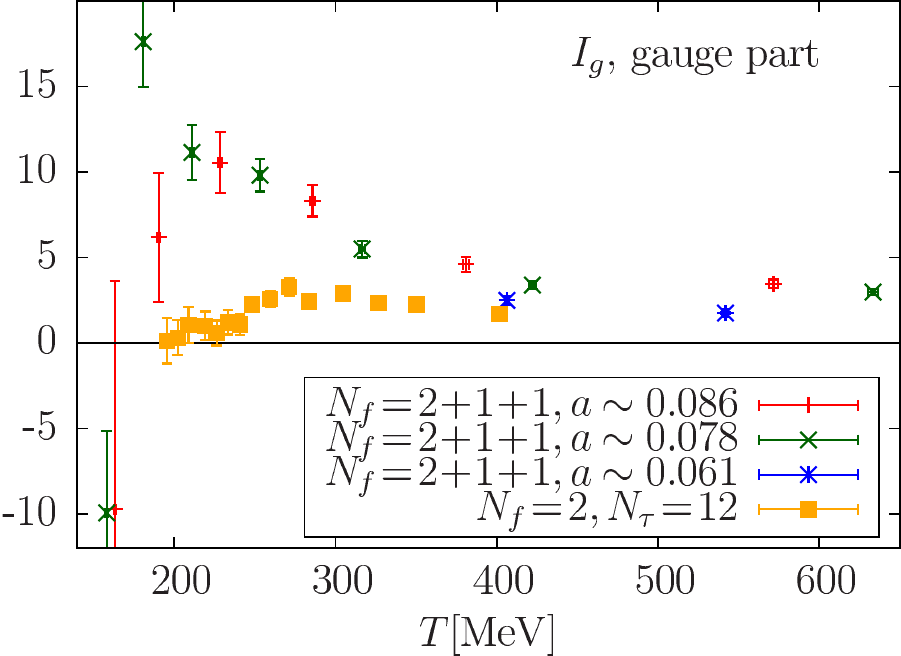} \hfill
}
 \caption[]{\links Fit of the $\beta$-function (\Eq{eq_fitfun}). \rechts Gauge part of the trace anomaly $I_g$ defined in \Eq{eq_traceanomaly_gauge}. For comparison we again show the 
 corresponding quantity in the two flavor case. }
\label{fig_traceanomaly}
\end{figure}

The standard approach that is taken in lattice calculations to evaluate the 
EoS as the temperature dependence of pressure $p(T)$ and 
energy density $\epsilon(T)$ begins with evaluating the trace anomaly 
\begin{equation}
  I = \epsilon - 3 p = - \frac{T}{V} \frac{d \ln Z}{d \ln a}  \;.
  \label{eq_traceanomaly}
\end{equation}
As $I/T^4$ is identical to a derivative of the reduced pressure $p/T^4$ with respect to the $\log$ of temperature
\begin{equation}
 \frac{I}{T^4} = T \frac{\partial}{\partial T} \left (  \frac{p}{T^4}  \right ) \;,
 \label{eq_pressurederiv}
\end{equation}
the pressure itself may be evaluated as an integral once $I(T)$ is known. The trace anomaly \Eq{eq_traceanomaly}
which constitutes a total derivative of the partition function with respect to the lattice spacing is calculable on the lattice by
taking the derivative of the lattice action with respect to the bare parameters first. The terms corresponding to the derivatives of various parts of the action have to be rendered 
finite by subtracting the corresponding expectation value at $T=0$. We denote the result by $\vevsub{\ldots}$ in 
the following. The lattice spacing dependence is encoded in the $\beta$-function as well as in additional functions encoding the running of the masses.
In what follows we restrict ourselves to the gauge parts of the trace anomaly which are most easily evaluated and make up its dominant part:
\begin{equation}
   I_g = \frac{\epsilon - 3 p}{T^4} (\mathrm{gauge~part}) = a \frac{d \beta}{d a} \vevsub{S_g} \, .
   \label{eq_traceanomaly_gauge}
\end{equation}
In order to evaluate the $\beta$-function $a \frac{d \beta}{d a}$ we adopt a similar strategy as in \Cite{Umeda:2012er} and perform a global fit 
of the coupling in terms of the rho meson mass ($m_\rho$) and the charged pion mass ($m_\pi$):
\begin{equation}
   \beta = c_0 \log{\left ( c_1 \left (a m_\rho \right ) \right )} + c_2 \left (a m_\rho \right )^2 +  c_3 \frac{\left ( a m_\rho \right )}{\left ( a m_\pi \right )}
   \label{eq_fitfun}
\end{equation}
For the determination of the rho mass we have used propagators that have been calculated for \cite{Burger:2013jya}.
We propagate the errors on the hadron masses to errors on the bare couplings $\beta$ by standard error propagation neglecting 
correlations and evaluate the value of $\chi^2$ per degree of freedom from these. We obtain a satisfying fit with $\chi^2/ \mathrm{dof} \approx 1.0$ that is 
shown in the left panel of \Fig{fig_traceanomaly}. The $\beta$-function is then evaluated using the rho mass to set the scale as 
\begin{equation}
   a \frac{d \beta}{d a} = (a m_\rho) \frac{d \beta}{d(a m_\rho)}
   \label{eq_evalbetafun}
\end{equation}
at the chosen simulation points (depicted in blue in the left panel of \Fig{fig_traceanomaly}). We show additionally the slopes $\frac{d \beta}{d(a m_\rho)}$ at all 
zero-temperature ETMC data points as thick black arrows. Thin grey arrows indicate the $\beta$-values the fit chooses for the different points.
In order to estimate the uncertainty connected with the choice of the fit function we have varied it by adding a term $\propto \left ( a m_\rho \right )^2 / \left ( a m_\pi \right )^2$ 
or by replacing the term $\propto \left (a m_\rho \right )^2$ with it.
The maximal magnitude of change in the $\beta$-function was then taken as a systematic error and included additionally to the statistical error of $I_g$. 
In the right panel of \Fig{fig_traceanomaly} the gauge part of the trace anomaly is shown together with the preliminary two-flavor result from \Cite{Burger:2012zz}.
As compared to the two-flavor case the peak in $I_g/T^4$ evaluated for $N_f = 2+1+1$ is significantly higher as well as shifted to smaller temperatures.  

\vspace{-0.3cm}
\section{Conclusions and Outlook}
We have presented first results of a fixed-scale finite-temperature study of the QCD crossover with a full dynamical second quark generation in the Wilson twisted
mass lattice regularization. As a first step we have concentrated on a single pion mass value together with three lattice spacings in order to estimate cutoff effects.
We find the latter to be small around the transition region for several observables, while at large temperatures they are significant. From the maximum in the 
disconnected part of the bare chiral susceptibility we have estimated the temperature of the chiral crossover. Comparing to the two-flavor case previously studied, we observe 
a downward shift of the transition temperature by approximately $20$ MeV. All studied observables show this behavior.

Furthermore, we have presented results for the gauge part of the trace anomaly. We are currently working on the inclusion of the various 
fermionic contributions and on the determination of the remaining mass related $\beta$-functions.
As the chosen fixed-scale approach can only provide a limited resolution in temperature, even more so if only even values of $N_\tau$ are considered,
we are working on including also odd numbers of temporal lattice sites.
Furthermore, we will address lower quark masses in the near future.

\vspace{-0.3cm}
\section*{Acknowledgements}
We are grateful to the CINECA and the HLRN supercomputing centers Berlin/Hannover as well as the FZ-J\"ulich
for providing computing resources used in this project. F.B., G.H.~and 
M.M.P.~acknowledge support by the DFG-funded corroborative research center SFB/TR 9. 
G.H. grate\-fully acknowledges the support of the German Academic National Foundation (Studienstiftung des deutschen Volkes e.V.) and of the
DFG-funded Graduate School GK 1504.
We thank A. Ammon for providing input for the bare 
strange Oster\-walder-Seiler mass.

\providecommand{\href}[2]{#2}\begingroup\raggedright\endgroup


\end{document}